\documentclass[11pt,a4paper]{article}
\usepackage{graphicx}
\usepackage{dcolumn}
\usepackage{bm}
\usepackage[english]{babel}
\usepackage{t1enc}
\usepackage{subfigure}
\usepackage{times}

\begin{document}
\title{Searching for signals of magnetic lensing in ultra-high energy cosmic rays}
\author{Geraldina Golup, Diego Harari, Silvia Mollerach and Esteban Roulet\\CONICET and Centro At\'omico Bariloche \\ Av. Bustillo 9500, 8400, S. C. de Bariloche, Argentina}
\maketitle
\begin{abstract}
Ultra-high energy cosmic rays are mostly charged particles and they are therefore deflected by magnetic fields on their path from their sources to Earth. An interesting phenomenon arising from these deflections is the appearance of multiple images of a source, i.e. cosmic rays with the same energy coming from the same source that can arrive to the Earth from different directions. In this work we present a technique to identify secondary images, produced by the regular component of the galactic magnetic field, benefiting from the fact that near caustics the flux is significantly magnified.
\end{abstract}

\section{Introduction}

Galactic and extragalactic magnetic fields do not only deflect the trajectories of charged cosmic rays but they can also amplify or demagnify the flux arriving from a source, modifying its spectrum. A detailed analysis of magnetic lensing effects can be found in refs. \cite{toes,sign,turb,spec}. An interesting phenomenon caused by magnetic lensing is the existence of multiple images, i.e. cosmic rays of the same energy coming from the same source that can arrive to the Earth from several different directions. The secondary images appear in pairs in the same direction of the sky at some critical energy $E_0$ which depends on the magnetic field model, the cosmic ray charge $Z$ and the direction to the source. When a pair of new images appear in the sky their flux is strongly magnified. For energies lower than the critical energy $E_0$, the arrival directions of cosmic rays of the secondary images move away from each other and their magnification decreases. Typically the flux of particles arriving to the Earth from the two secondary images in the energy range [$0.9 E_0, E_0$] is an order of magnitude larger than the flux that would arrive if no lensing effects were present. Therefore a clustering of events with similar energy is expected around the direction where the secondary images appear and it is thus possible to benefit from this fact and design a method to search for such signals in cosmic ray data as proposed in this paper.\\

In analogy with the gravitational lensing effect, we call caustics the positions of the sources for which a pair of images with diverging amplification appear. These form closed lines in the sky whose location is a function of the energy. The corresponding positions of the pairs of new images are the critical lines. Their shape and position as a function of the energy are characteristic of each galactic magnetic field model. The identification of the direction and energy at which the new images appear would then provide new insights on the galactic magnetic field. Another handle on magnetic fields can be obtained from the observation of ultra-high energy cosmic ray multiplets (events with different energy that come from a single point-like source), which could be used to determine the integral of the perpendicular component of the galactic magnetic field along the cosmic ray trajectory \cite{multipletes}.\\

We note that for the galactic magnetic field model we consider the first caustics appear at energies $E_0/Z \approx 30$ EeV. If the cosmic ray composition was heavy, the lensing effects we study will only appear at energies above the maximum energies detected by existing experiments. Hence, our results essentially apply to proton cosmic ray sources. In addition, even in the case in which the source produces heavy nuclei with a unique value of $Z$, photodisintegration during propagation will lead to a distribution of values of the charge reaching Earth, and the caustics will appear at different energies for each charge, diluting the possible signals. Effects of magnetic deflections of heavy nuclei in galactic magnetic fields have been recently discussed in \cite{giacinti}.\\

The paper is structured as follows: in section 2 we describe the general known characteristics of the galactic magnetic field and review some magnetic lensing facts. In section 3, we explain the method used to search for signals of multiple images of a source and to characterize them. In section 4, we apply the method to simulations and estimate the density of sources for which the effect is expected to be observable. Finally, section 5 is a summary of the results and contains our conclusions.\\

\section{Galactic Magnetic Field and Magnetic Lensing}

Although there are several observational methods to obtain information about the galactic magnetic field (Zeeman splitting, polarized thermal emission from dust in clouds, polarization of starlight, synchrotron radio emission and Faraday rotation of polarized sources \cite{han1,beck,brown}), this field is not yet precisely known.\\

The galactic magnetic field has a large-scale regular component and also a turbulent one. From the analysis of the polarization of starlight it is known that the local regular component is approximately parallel to the galactic plane and follows the local spiral arms \cite{heiles}. The regular component has a local value between $B_{reg}\simeq 2$-$4\ \mu$G according to Faraday rotation measures \cite{han3} or synchrotron measurements \cite{beck} (the possible origin of this difference is analyzed in \cite{beck2,cowsik,sarkar}). Reversals in the direction of the magnetic field between neighboring arms have been detected with Faraday rotation data, which provide evidence that in the inner arm magnetic fields are counterclockwise when viewed from the North Galactic pole, while in the local region the fields in the interarm regions are clockwise \cite{han1}. The galactic halo field remains a puzzling subject since rotation measures indicate that it appears to be antisymmetric with respect to the galactic plane \cite{han1}, but it is not yet understood how can a halo field with odd parity be superimposed with a disk field with even parity \cite{beck}. The turbulent component has a root mean square amplitude of $B_{rms} \simeq (1$-$2) B_{reg}$ and a typical coherence length $L_c\simeq$100 pc \cite{rand, ohno}. The dominant effect on the deflection of high energy charged particles traveling through the Galaxy is produced by the regular component because it is coherent on scales much larger than $L_c$.\\

Even though the presence of extragalactic magnetic fields affects the propagation of cosmic rays, they are probably not the predominant effect in the deflection of ultra-high energy charged particles. For instance, from the results in \cite{dolag} one finds that for 10 EeV protons the deflections are smaller than 1 degree in $98 \%$ ($90 \%$) of the sky for distances to the source of 35 Mpc (70 Mpc). However, the strength of the extragalactic magnetic fields is quite uncertain and for example in \cite{arme} larger values for the deflections are found. We will only consider hereafter the deflection produced by the galactic magnetic field.\\

\begin{figure}[!t]
\begin{center}
\subfigure[\label{fig1a}]{\includegraphics[scale=0.65]{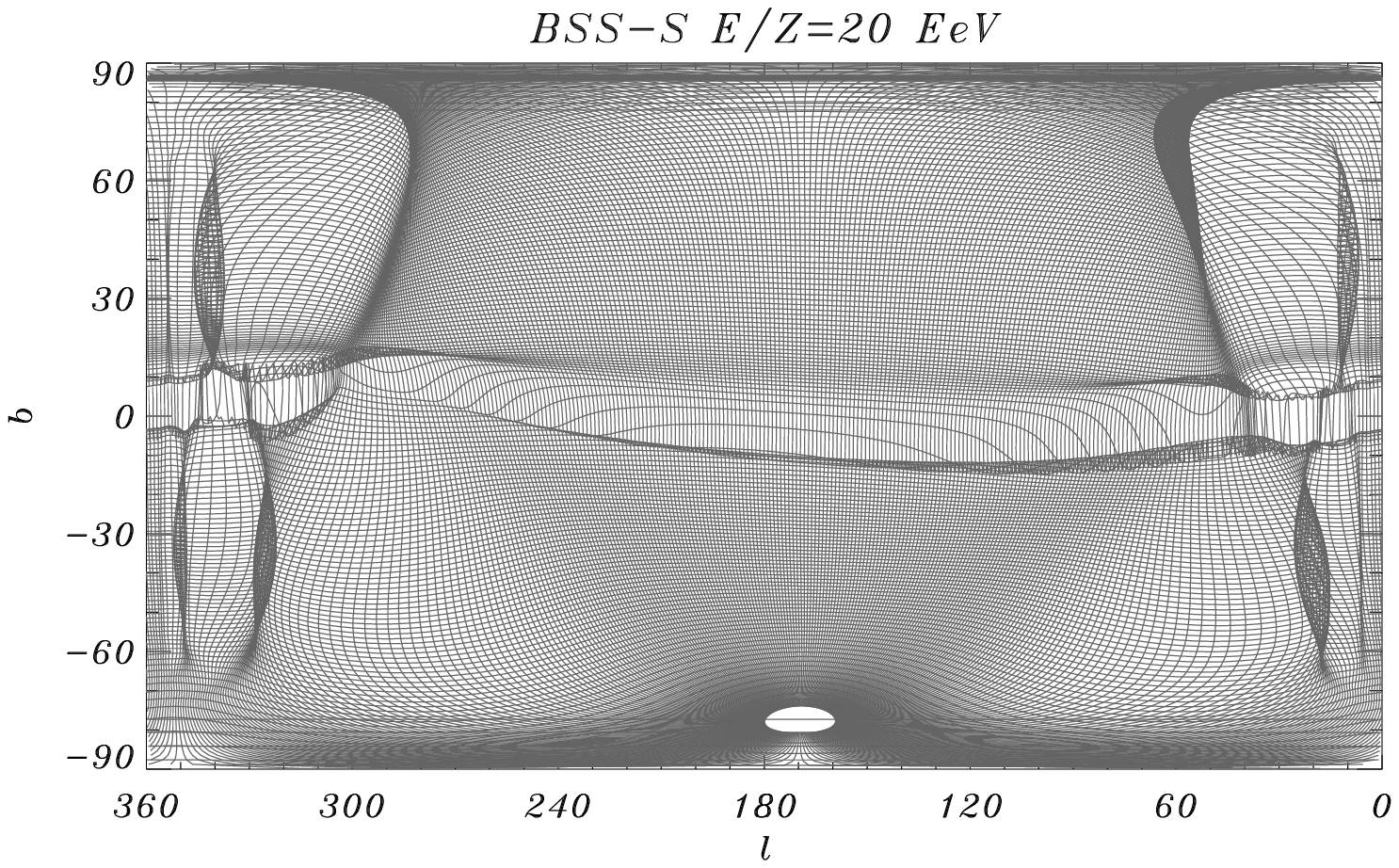}}
\subfigure[\label{fig1b}]{\includegraphics[scale=0.65]{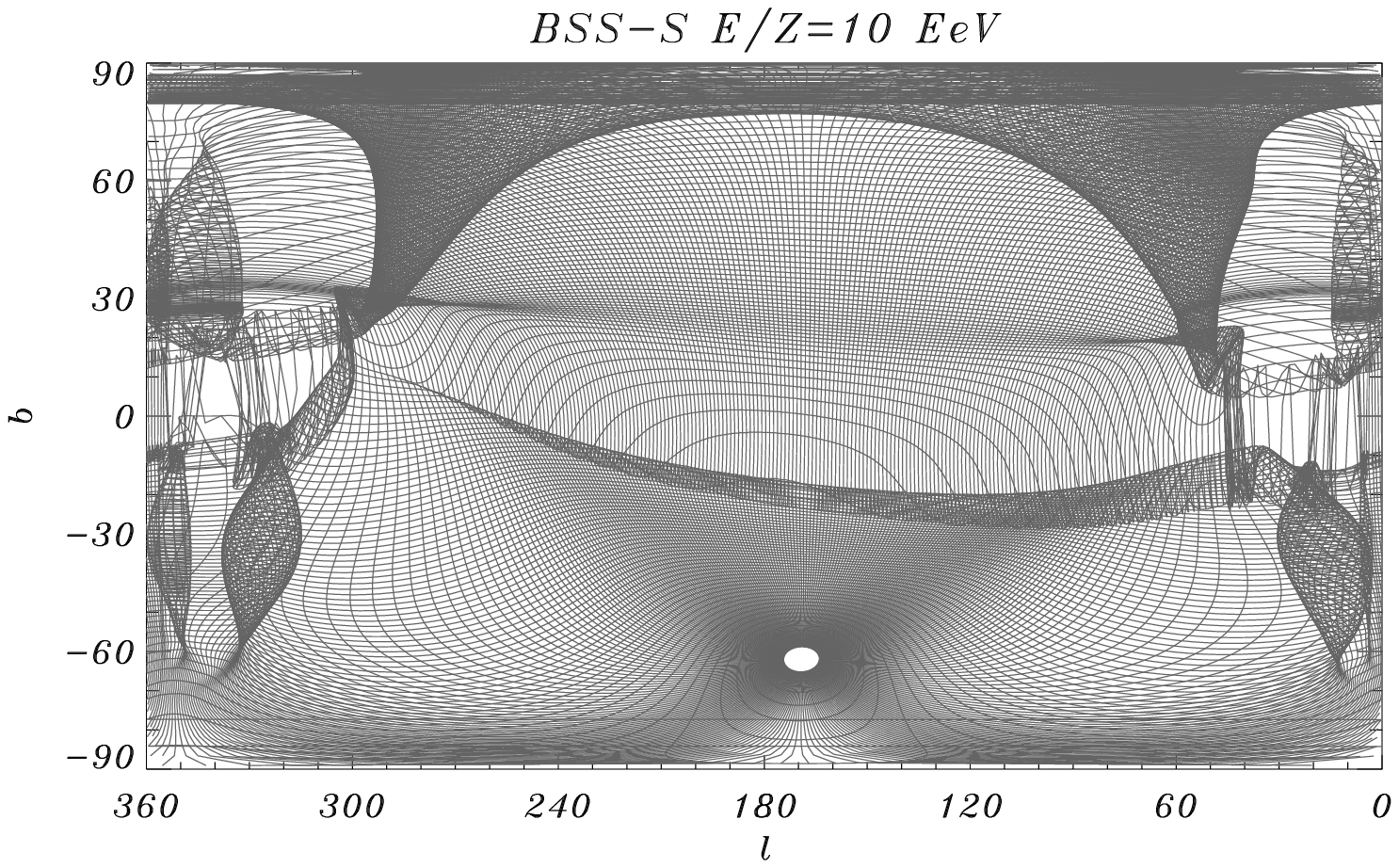}} \caption[]{`Sky sheets': directions of
incoming cosmic rays in the halo that correspond to a regular grid of arrival directions at Earth, adopting the BSS-S
magnetic field configuration, for particles with $E/Z$ = 20 EeV (a) and 10 EeV (b).}\label{fig1}
\end{center}
\end{figure}

In this paper we consider an illustrative galactic field model that includes some of the observed features, such as field reversals, and reproduces the local field strength. The regular component of the galactic magnetic field is modeled with a bisymmetric field with even symmetry (BSS-S) with structure and strength very similar to those used in \cite{stanev} but smoothed out as described in \cite{toes}. In this model the galactic magnetic field reverses its sign between the arms of the Galaxy and the field is symmetric with respect to the Galaxy's mid-plane. The local value of the field is taken as $2\ \mu$G. For the dependence on $z$ (i.e. perpendicular to the galactic plane) a contribution coming from the galactic disk and another one from the halo are considered:
\begin{equation}
\vec B_{reg}(x,y,z)=\vec B_{reg}(x,y,z=0) \left(\frac{1}{2\cosh(z/z_1)}+\frac{1}{2\cosh(z/z_2)}\right),
\end{equation}
with $z_1$= 0.3 kpc and $z_2$= 4 kpc.\\

\begin{figure}[!t]
\begin{center}
\subfigure[]{\includegraphics[scale=0.40]{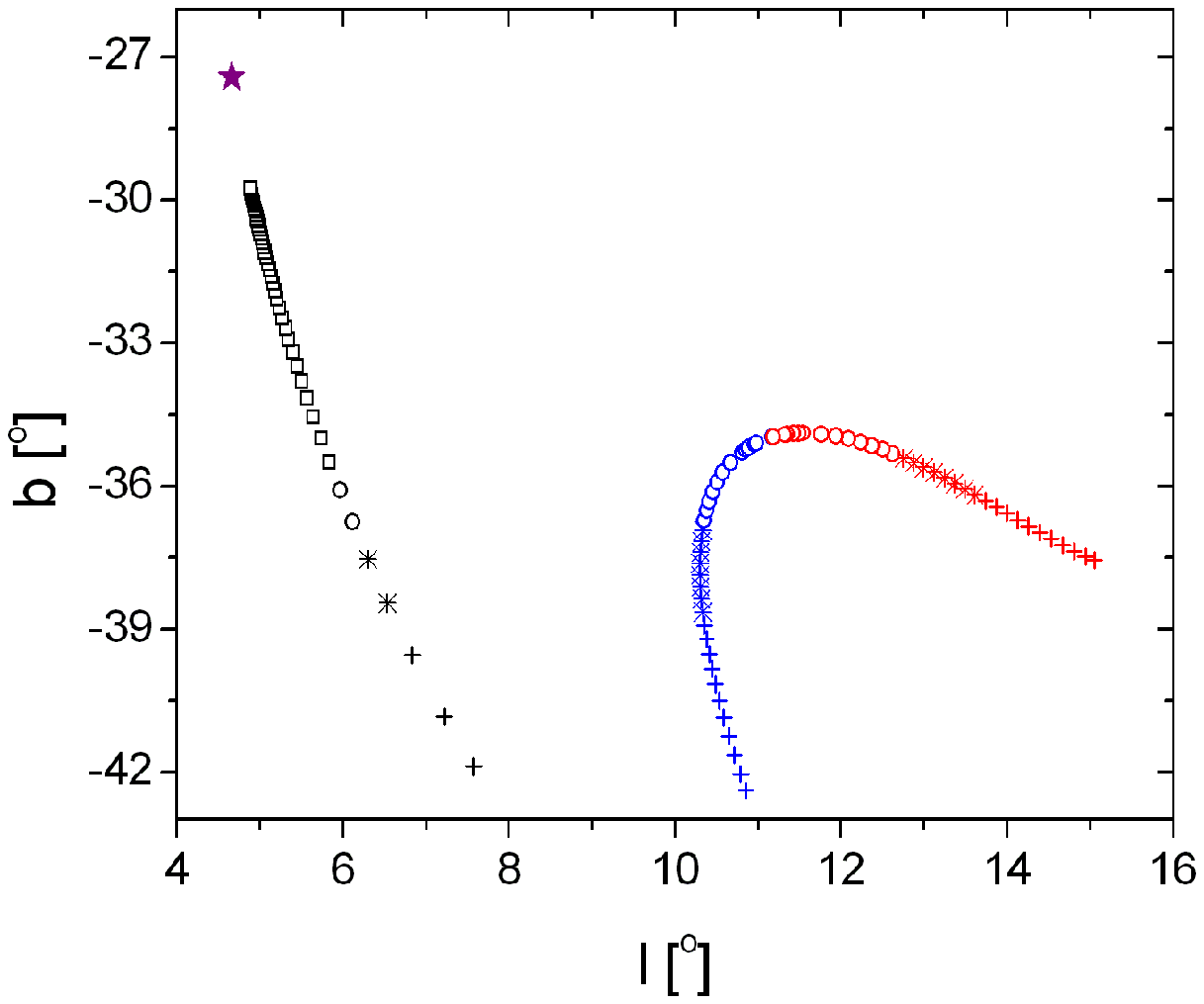}}
\subfigure[]{\includegraphics[scale=0.40]{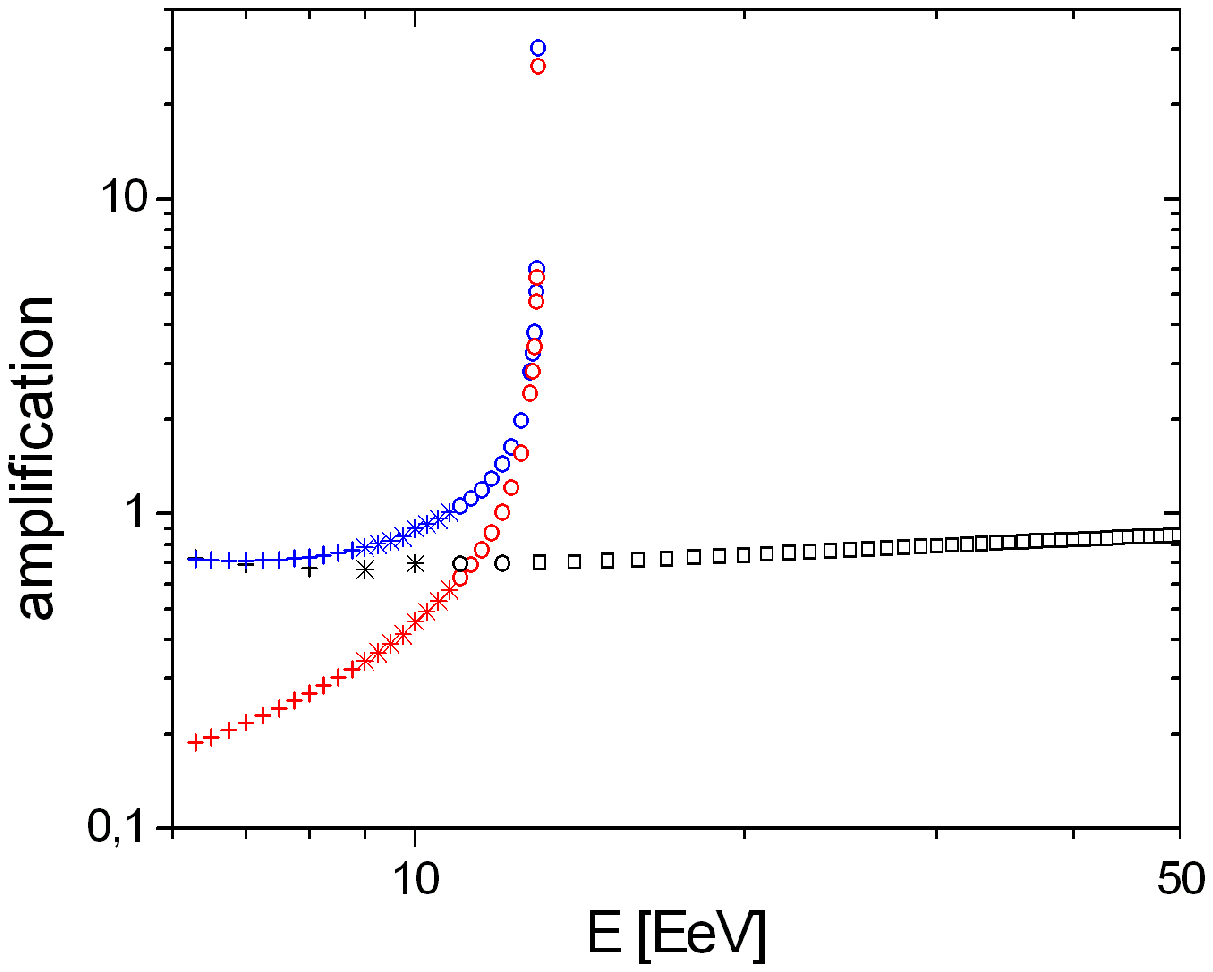}} \caption[]{Position of the images in galactic coordinates for a source located at ($l,b$)=($4.7^\circ,-27.4^\circ$)(a) and their amplifications as a function of the energy (b). Black symbols correspond to the main image while blue and red symbols correspond to secondary images. A star indicates the position of the source. Plus signs correspond to events with energies between 6 and 9 EeV, asterisks correspond to energies between 9 and 11 EeV, circles correspond to energies between 11 and 13 EeV and squares correspond to energies above 13 EeV.} \label{fig2}
\end{center}
\end{figure}

The deflections caused by the magnetic field can lead to lensing phenomena, such as the energy dependent
magnification and demagnification of the flux that modifies the energy spectrum of the source, or to the
appearance of secondary images \cite{toes,sign,turb}. The magnitude of the flux amplification depends on the
arrival direction, on the ratio between the energy and charge $E/Z$ of the cosmic ray and on the magnetic field
model. The formation of multiple images and the flux magnification can be understood pictorially plotting for a
regular grid of arrival directions at Earth the corresponding directions from which the particles arrived to the
galactic halo. This is shown in Figure \ref{fig1} for the model of the regular magnetic field configuration considered and for particles with $E/Z = 20$ EeV (top panel) and 10 EeV (bottom panel), where 1
EeV $\equiv 10^{18}$ eV. One may picture this distorted image of the sky seen from the Earth as a sheet (the
`sky sheet') that can be stretched and folded. A source located within a fold will have multiple
images, i.e. cosmic rays of the same energy can arrive from this source to the Earth along several different directions. At each energy the position of the folds in the plots indicate the location of caustics. If a caustic crosses the position of a source at an energy $E_0$, a pair of new images appears and their magnification diverges as $1/\sqrt{E_0 - E}$ for energies close to $E_0$ \cite{sign}. Therefore, a clustering of events with energies close to $E_0$ is expected, increasing the probability of detecting the multiple images. As the flux magnification depends on the observed direction, at lower energies each secondary image will have different amplification. Pictorially speaking, the flux coming from a source in a region where the sheet is stretched will appear demagnified while that from a source in a compressed region will appear magnified. As the sky sheet changes with the energy, the spectrum observed from a given source is different from the emitted one \cite{toes}. In Figure \ref{fig2} we show for illustration the position of the images and their amplifications as a function of the energy for a source located at galactic coordinates ($l,b$)=($4.7^\circ,-27.4^\circ$). In this example, one has $E_0/Z=13.0$ EeV.\\

\begin{figure}[!ht]
\begin{center}
\includegraphics[scale=0.65]{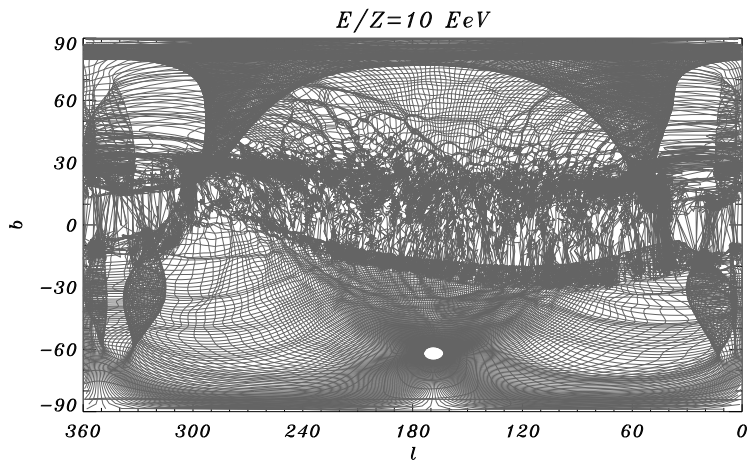}
\caption[]{`Sky sheet' adopting the BSS-S magnetic field configuration and including a turbulent component, for particles with $E/Z$ = 10 EeV. The turbulent component was simulated with a Gaussian random field with zero mean and root mean square amplitude of $B_{rms}=2 \mu$G and a coherence length of $L_c=100$ pc.} \label{fig3}
\end{center}
\end{figure}

\begin{figure}[!ht]
\begin{center}
\subfigure[\label{fig4a}]{\includegraphics[scale=0.60]{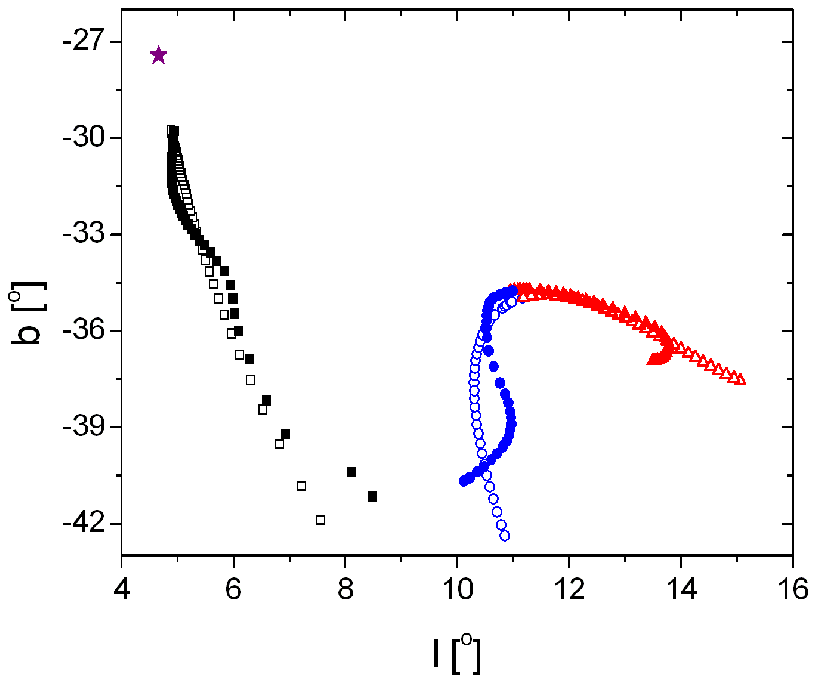}}
\subfigure[\label{fig4b}]{\includegraphics[scale=0.60]{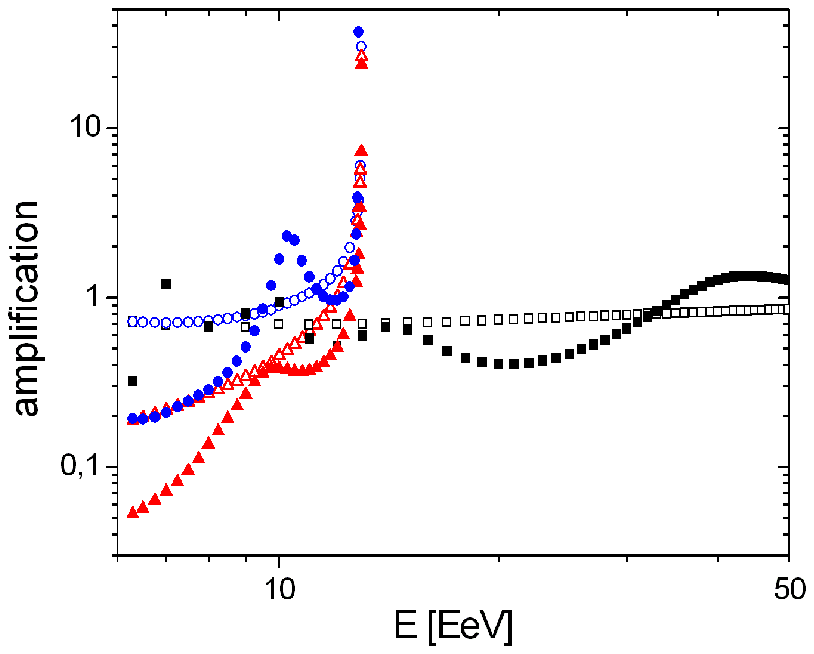}} \caption[]{Position of the images in galactic coordinates for a source located at ($l,b$)=($4.7^\circ,-27.4^\circ$) (same as the example in Figure 2) (a) and their amplifications as a function of the energy (b). Full symbols are used for events propagated through both the regular and turbulent component, while empty symbols are used to indicate the events which have only been propagated through the regular field.}\label{fig4}
\end{center}
\end{figure}

Magnetic lensing phenomena also appear for turbulent fields \cite{turb}. In this case, multiple images appear
below a critical energy $E_c$ such that the typical transverse displacements among different paths after traveling
a distance $L$ in the turbulent field become of the order of the correlation length of the random magnetic field
($\delta_{rms} \sim L_c/L$). This critical energy is typically higher than the corresponding one for the regular
magnetic field, but the folds produced by the random field near the critical energy are on a much smaller
angular scale. For decreasing energies, the fraction of the sky covered with folds increases, however the
magnification peaks become increasingly narrower and for $E<E_{c}/3$ their integrated effect becomes less
noticeable. This can be pictured in terms of the `sky sheet' (Figure \ref{fig3}), where small folds due to the turbulent component appear on top of the large folds produced by the regular component, being hence these last ones those giving the dominant effects. In Figure \ref{fig4}, we plot the same example as in Figure \ref{fig2}, but with the events propagated through both regular and turbulent components of the galactic magnetic field. We see that the main properties of the magnification peaks which are relevant for our analysis (their location, $E_0$ and their approximate shape within $10\%$ of $E_0$, as well as the elongated shape of the images near the caustic) remain essentially unchanged. Hence, on the following we will focus on the effects of the regular field alone.\\

Let us note that a possible background to the signals due to the appearance of multiple images (i.e. a clustering of events with similar energies) could be produced in bursting models for the sources of ultra-high energy cosmic rays. Due to the energy dependent delay in the arrival time of cosmic rays in the presence of intergalactic magnetic fields, this would imply that at a given time only cosmic rays with the same time delay, and therefore same energy, would be observed at Earth coming from a bursting source \cite{waxman}. However, the lensing features should appear only below some critical energy (at which the first caustics appear in the sky) while those due to bursting sources should be relevant up to the highest energies.\\

\section{Method}

\begin{figure}[!t]
\begin{center}
\subfigure[\label{fig5a}]{\includegraphics[scale=0.65]{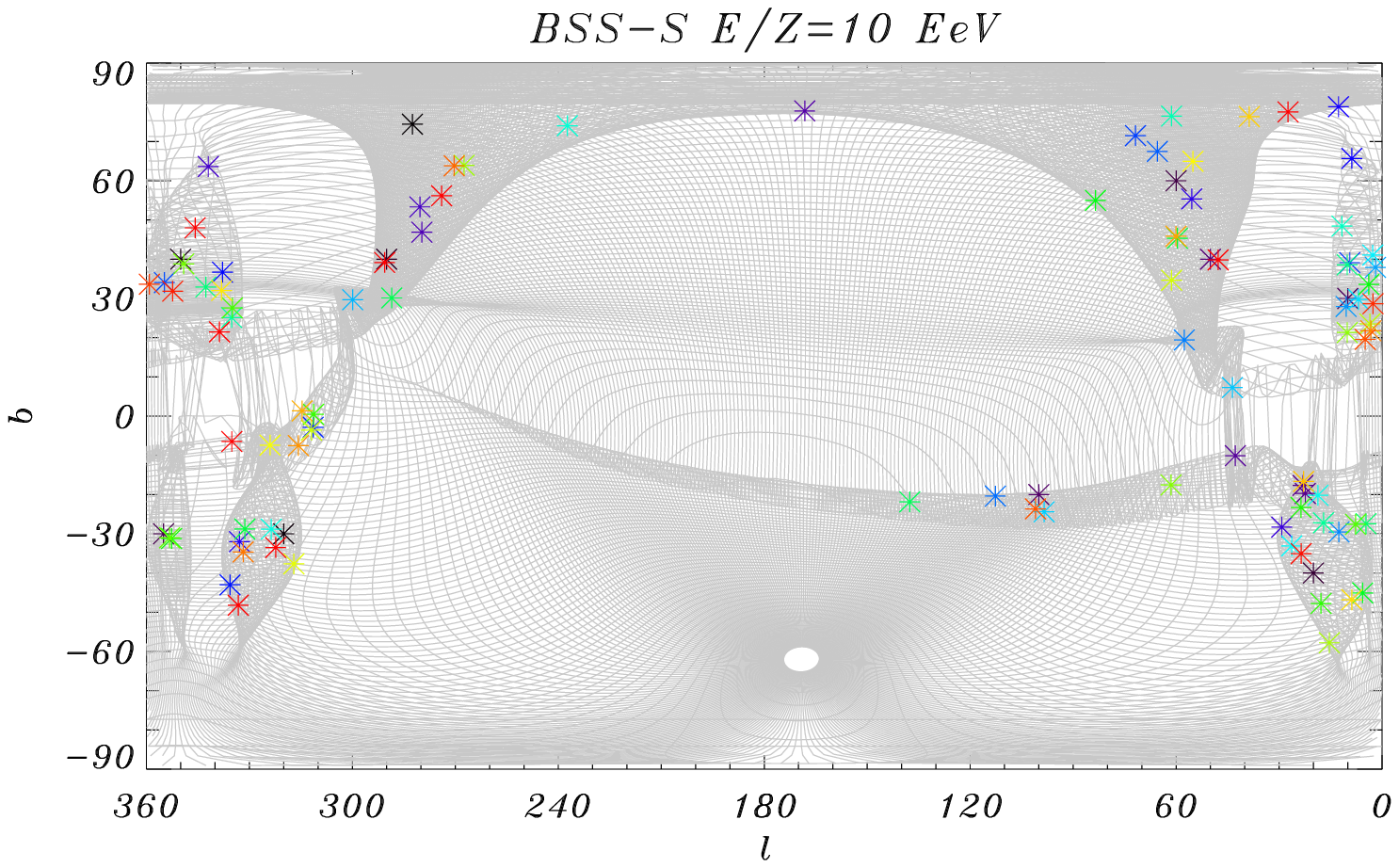}}
\subfigure[\label{fig5b}]{\includegraphics[scale=0.65]{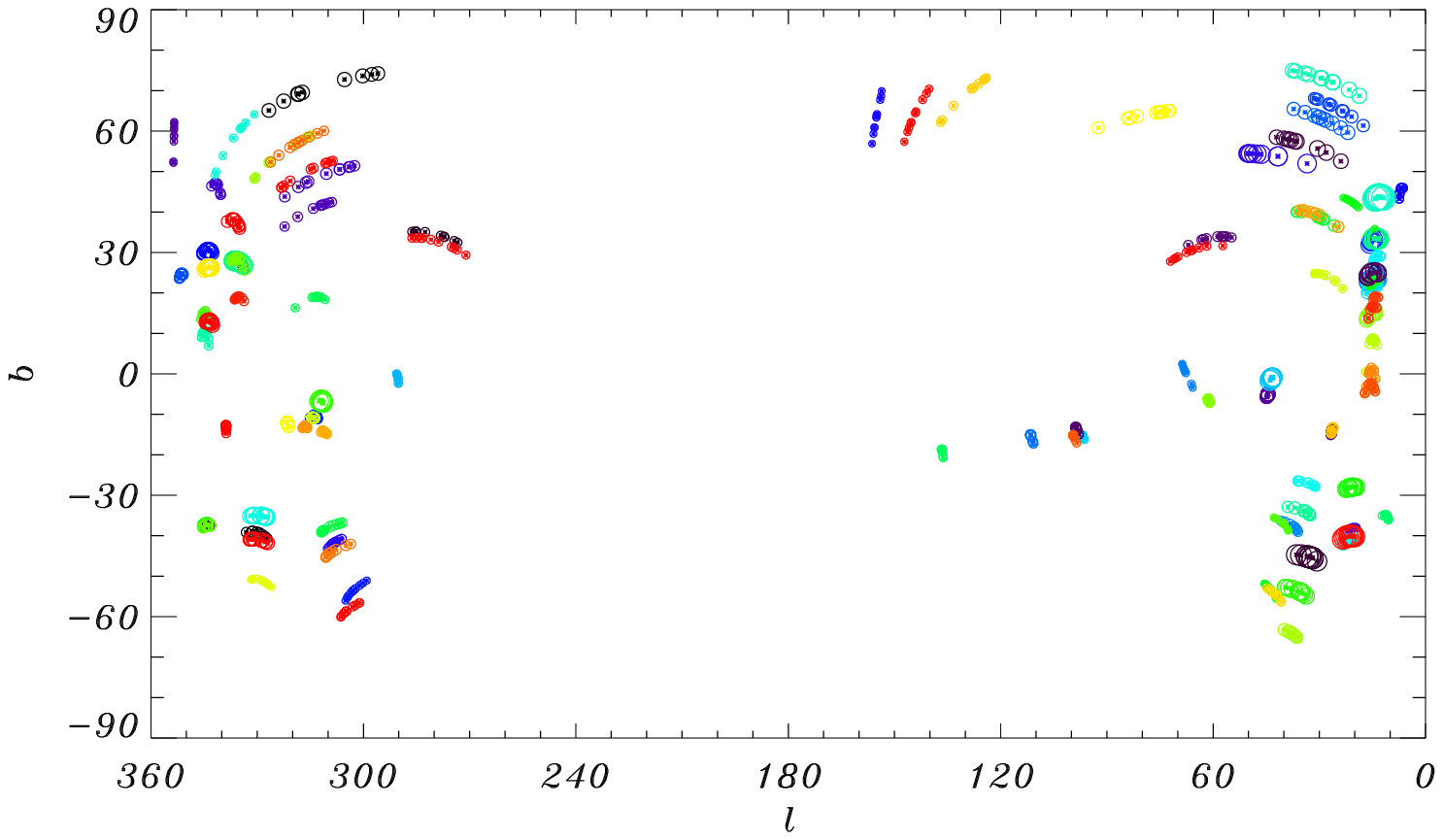}}
\caption[]{One hundred simulated sources (asterisks) that have multiple images in a projection of the celestial sphere in galactic coordinates overlaid with a `sky sheet' of the BSS-S galactic magnetic field model for particles with $E/Z$ = 10 EeV (a) and ten events from secondary images from each of these sources (b). The radius of the circles is proportional to the energy of the events.}
\end{center}
\end{figure}

The purpose of this work is to design a method to search for signals of multiple images of a source, taking into account that cosmic rays from that source should appear clustered in the sky at energies close to the critical one. In order to define what would be the typical signal of the clustering due to the appearance of multiple images of an extragalactic point-like source, we simulate sources that are randomly distributed in the sky and keep those that are crossed by caustics for energies above a given threshold. We note that for the BSS-S magnetic field model that we adopted, sources located in approximately 24$\%$ of the sky would have caustics at energies above 10 EeV and assuming a proton composition, while above 20 EeV this fraction is only 7$\%$. In Figure \ref{fig5a} we show the location of the one hundred randomly located sources from which we simulate the events which are shown in Figure \ref{fig5b}.\\

For each source we simulate cosmic rays which we then propagate through the galactic magnetic field. These simulations are done backtracking antiprotons leaving the Earth up to a distance where the effect of the galactic magnetic field becomes negligible \cite{toes}. The initial direction of the backtracked antiproton (that will be associated to the arrival direction of protons of the same energy) is recursively adjusted until the final direction points to the source with an accuracy better than $10^{-4}$ degrees.\\

We adopt an $E^{-2}$ power spectrum at the source which is then modified when observed at the Earth by the magnification of the flux as a function of energy in the corresponding direction of the sky. We display in Figure \ref{fig5b} ten events from the secondary images of each source, with energies between 0.9 $E_0$ and $E_0$. The energy interval is chosen so as to take into account a reasonable experimental energy resolution, which is larger than the width of the amplification peak (defined as the energy above which the amplification is larger than 3) that is equal or smaller than 0.02 $E_0$ for $50 \%$ of the simulations. Note that being these magnification peaks so narrow in comparison with the typical experimental energy resolution, it is unfeasible to study detailed correlation between the location and the energy of the events appearing in these magnification peaks. The arrival directions of cosmic rays from both secondary images appear along a line in the sky since for each image the arrival directions move in opposite directions as the energy decreases below $E_0$. Then, this allows to measure the `width' and `length' (shortest and longest distance) of the distribution of the events in the sky. In order to study the effect of the angular resolution, a random gaussian deflection with standard deviation $\sigma=1^\circ$ (which is the typical resolution of air shower arrays at ultra-high energies) is added to the position of each event.\\

\begin{figure}[!t]
\begin{center}
\subfigure[\label{fig6a}]{\includegraphics[scale=0.40]{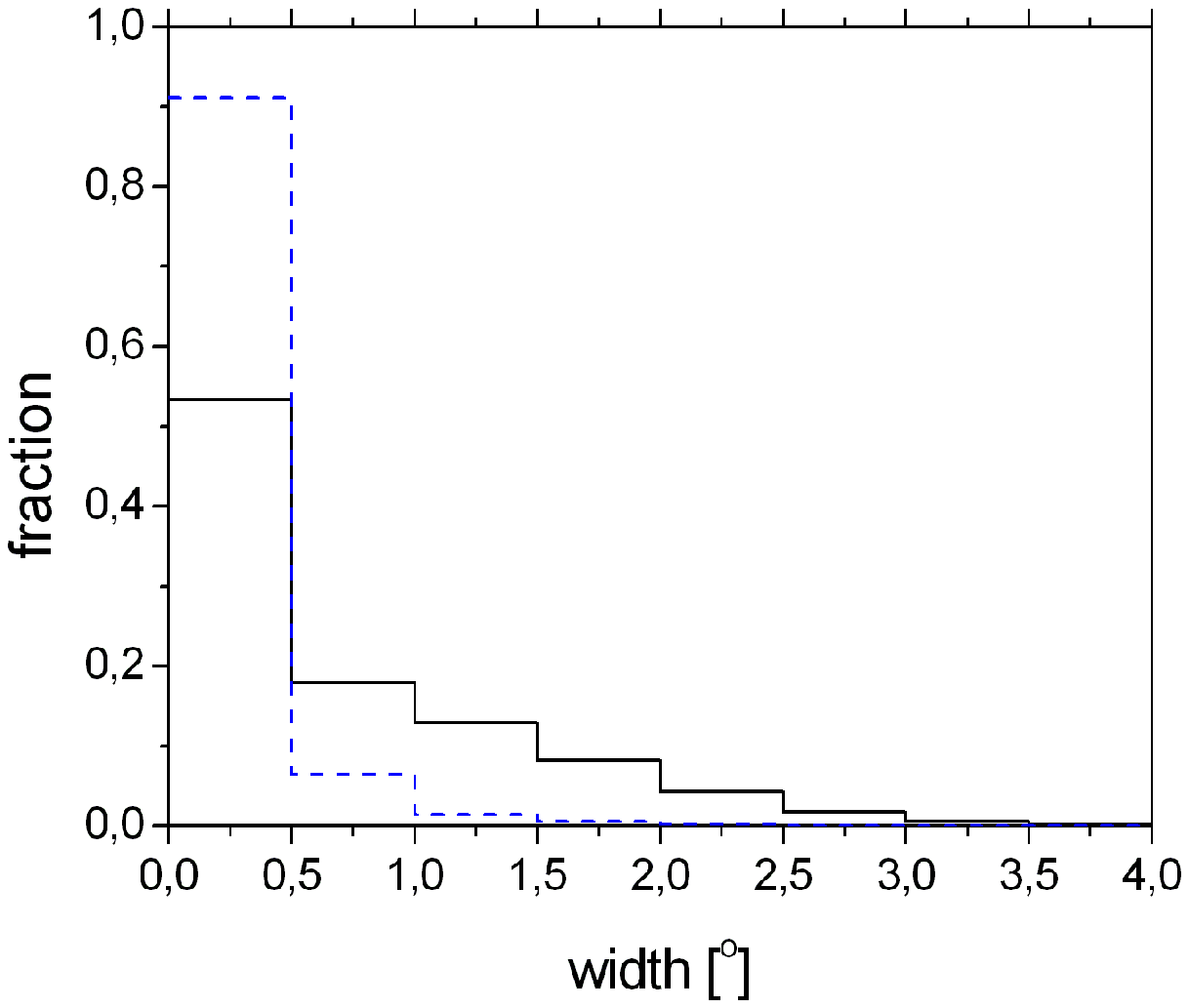}}
\subfigure[\label{fig6b}]{\includegraphics[scale=0.40]{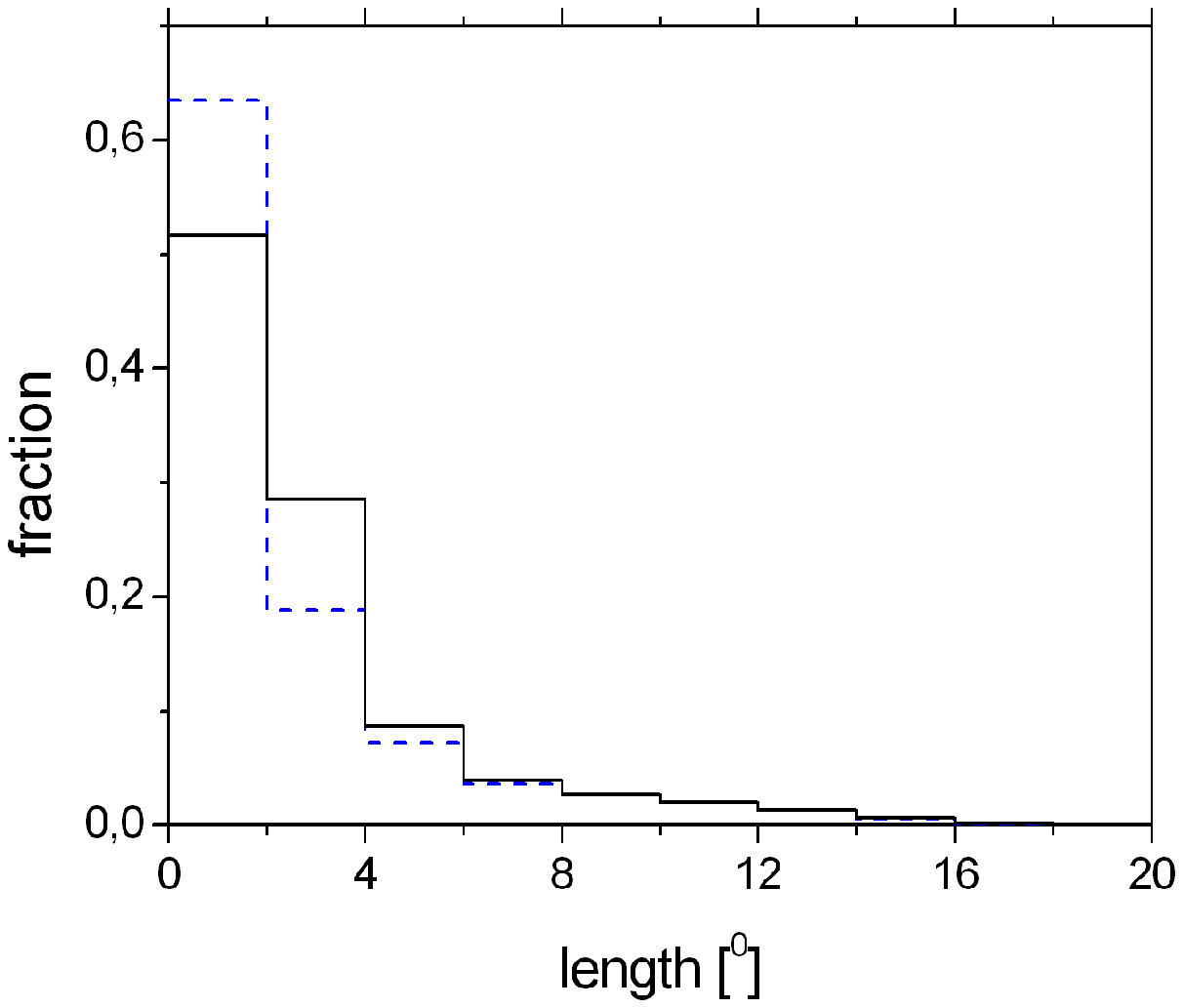}} \caption[]{Distribution of events within a certain `width' and `length' in the tangent plane of the secondary images with energies between [0.9 $E_0$, $E_0$] for the one hundred simulated sources, with (black solid line) and without (blue dashed line) angular resolution effects (of 1 degree standard deviation).}\label{fig6}
\end{center}
\end{figure}

In Figure \ref{fig6} we show the distribution of events within a certain width or length for sets of ten events from the one hundred simulated sources without and with angular resolution effects, performing in each case $10^4$ different realizations of the arrival directions and energies of the events. As the typical width without experimental uncertainty is below $1^\circ$, the inclusion of the experimental angular uncertainty has a significant impact on this quantity, while in the case of the length, which for some sources is as large as $18^\circ$, the angular resolution effects have only a minor impact. From these distributions we obtain the characteristic angular dimensions of the elongated excesses of events which would be seen in the sky due to the appearance of multiple images of a source. We use this to determine the angular window that should be adopted to perform the search of excesses due to multiple imaging. We will consider a maximum width of $2^\circ$, for which $92.5 \%$ of the simulated events are included (taking into account the angular resolution). Concerning the maximum length, we set it at $8^\circ$, for which $92.8 \%$ of the simulated events are contained. There is of course a compromise between a larger maximum length (or width), which will include the full length of the cosmic rays directions with energies between [$0.9 E_0, E_0$] for a larger fraction of simulations, and the increase of the number of events expected from the background. We note that the longer structures are limited to certain regions of the northern folds near the galactic center. On the other hand, restricting the search to structures of shorter length would just imply loosing some events with energies close to 0.9 $E_0$ and typically belonging to the least amplified secondary image. We note that $86.5\%$ of all the simulated events are located within angular windows of $8^\circ$ length and $2^\circ$ width.\\

Having characterized the typical signal of clustering due to the appearance of multiple images of point-like sources,
we construct an algorithm to search for this kind of features in cosmic ray data. The idea is to scan the sky counting the number of events with energies in a range [$0.9 E, E$], within an angular window of $2^\circ \times 8^\circ$ arbitrarily oriented and compare the result with the expected number for an isotropic distribution of events, as a function of $E$. The significance of the excess is calculated with the Li \& Ma method \cite{lima}. For the simulations done in this paper, we will consider a full-sky observatory with equal exposure in all directions. Real observatories have an exposure which is not uniform on the celestial sphere and this fact can be easily taken into account and should not affect our results significantly. Due to the scan performed both in the location and orientation of the angular window as well as on the energy, in order to compute the chance probability of finding an excess with a given Li \& Ma significance we just apply the same algorithm to simulations of isotropically distributed events to find what fraction of the isotropic simulations shows more clustering than the data.\\

\section{Application to simulations}

\begin{figure}[!t]
\begin{center}
\subfigure[\label{fig7a}]{\includegraphics[scale=0.40]{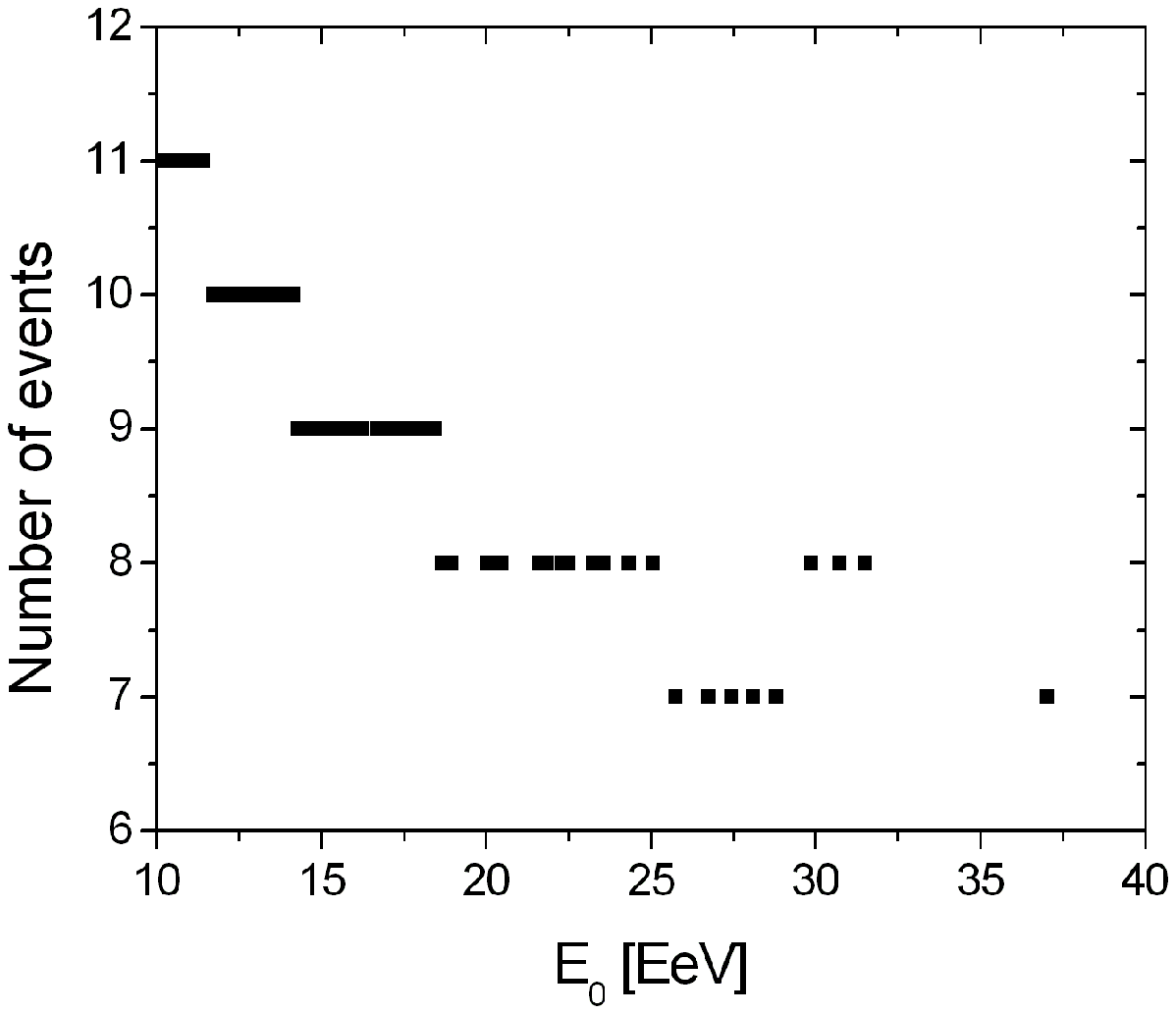}}
\subfigure[\label{fig7b}]{\includegraphics[scale=0.40]{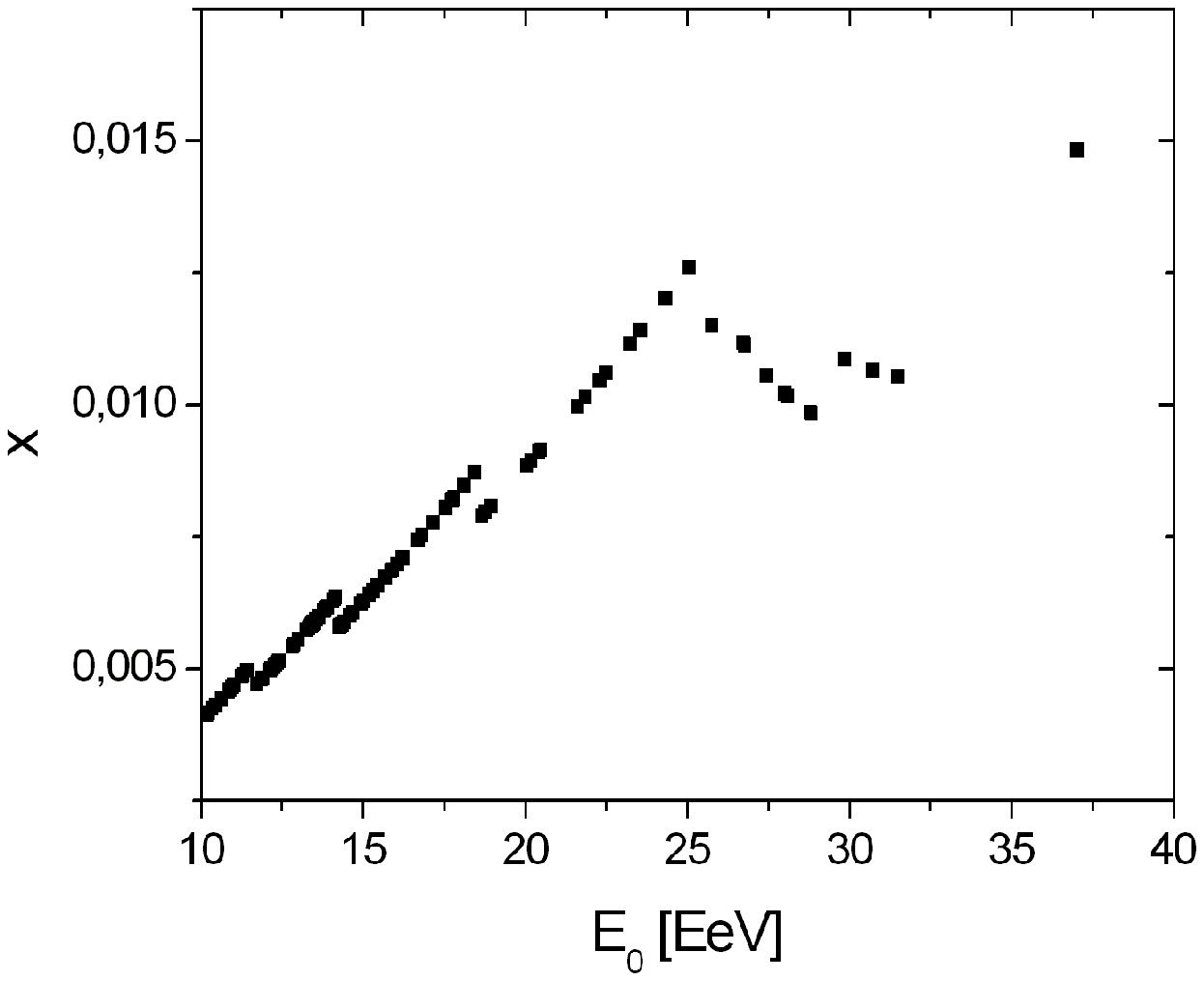}}
\caption[]{Number of events with energies between [0.9 $E_0$, $E_0$] that are necessary to have an excess with isotropic chance probability smaller than $10^{-3}$, as a function of the energy $E_0$ at which the multiple images appear (a) and the corresponding fraction $x$ of the cosmic ray flux in that energy range (b).}
\end{center}
\end{figure}

In this section we apply the method previously discussed to the simulations of sources with multiple images. To
begin with, we analyze how many events an excess would need to have in order that the probability for it to happen by
chance be smaller than $10^{-3}$. This is determined by analyzing simulations of isotropically distributed background events superimposed with events simulated from each of the sources with secondary images discussed in the previous section. The energy of the background events is selected following a broken power-law spectrum $E^{-\gamma}$, for which we adopt the values determined by the Pierre Auger Observatory \cite{espectro}, $\gamma=2.59$ for 4 EeV $\leq E \leq$ 28.8 EeV and $\gamma=4.3$ for $E >$ 28.8 EeV. For each of the simulated sources, the secondary images appear at a different rigidity $E_0/Z$ and due to the falling cosmic ray spectrum, the number of background events depends on energy (and on the total number $N_{tot}$ of cosmic rays measured above the threshold). Therefore, the number of events from a source required to have a detectable excess depends on $E_0/Z$ and $N_{tot}$. We present the results assuming the data set consists of $N_{tot}=10^4$ measured cosmic rays above 9 EeV, comparable to the present data set of the largest existing air shower array, the Pierre Auger Observatory. The number of events with energy in the range [$0.9 E_0, E_0$] and in an angular window of $2^\circ \times 8^\circ$ coming from each of the one hundred simulated sources that are needed in order that the probability that an equal or larger excess appears by chance be $10^{-3}$ is plotted in Figure \ref{fig7a} as a function of the energy of the caustic assuming proton composition of the source ($Z=1$). The number of required events increases as the energy decreases (from 7 for $E_0 \approx 20$ EeV to 11 for $E_0 \approx 10$ EeV), which is due to the fact that the background signal grows at lower energies. The slight increase at energies above 29 EeV is due to the change in the slope of the broken power law spectrum. We note that if the total number of ultra-high energy cosmic rays is increased, the number of events required to identify a secondary image with the specified significance will scale as $\sqrt{N_{tot}/10^4}$.\\

Measuring such an excess would allow to identify a direction and an energy at which a caustic crosses the position of a source. This information would be valuable in distinguishing between different galactic magnetic field models which lead to different predictions for the location of caustics.\\

The possibility to detect the signal of secondary images depends on whether there is a strong enough source of cosmic rays in a region crossed by a caustic so as to produce the required number of events. This is directly related to the density of cosmic ray sources in our neighborhood since a larger source density implies fewer expected events from each source, what would make less probable the detection of this effect.\\

\begin{figure}[!t]
\begin{center}
\subfigure[\label{fig8a}]{\includegraphics[scale=0.40]{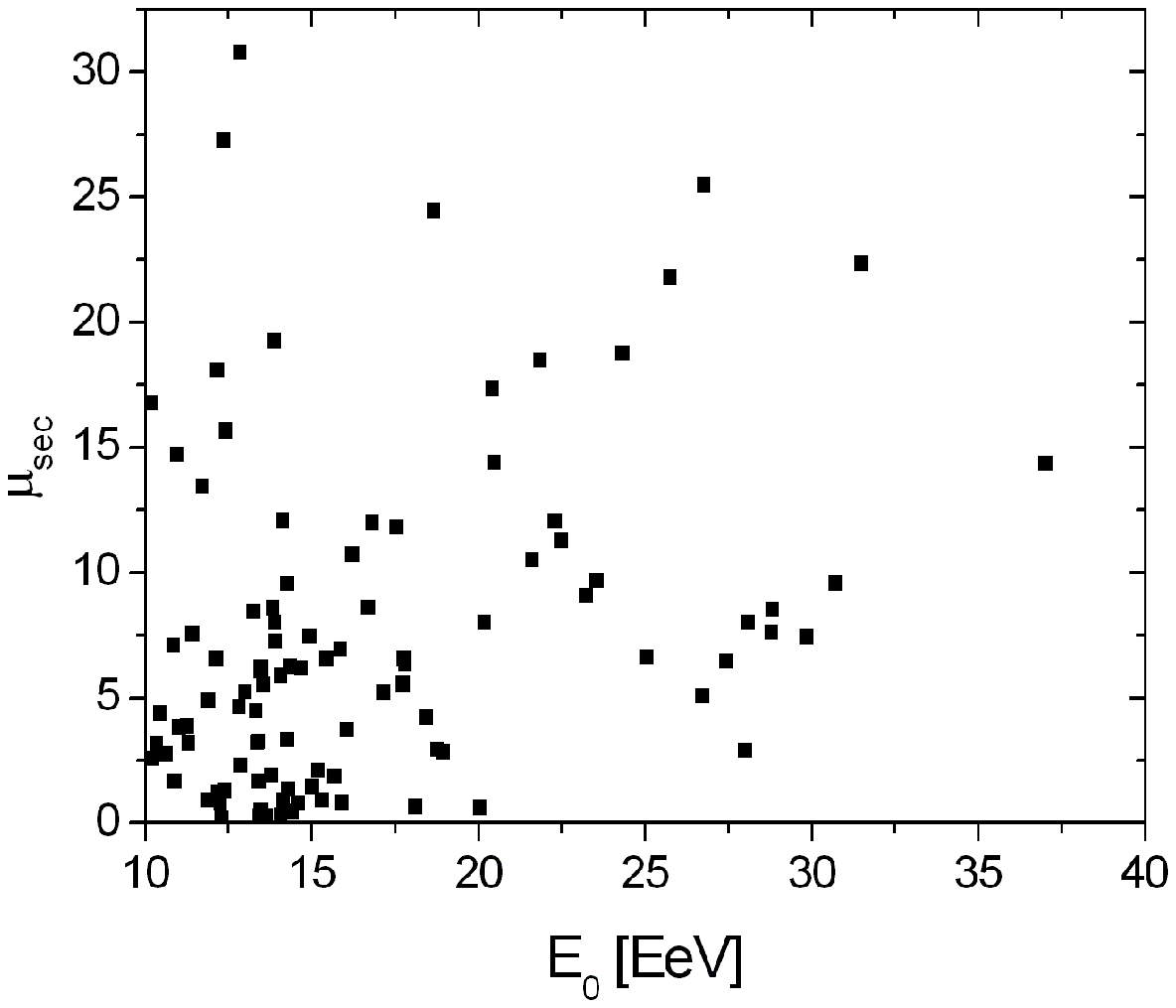}}
\subfigure[\label{fig8b}]{\includegraphics[scale=0.40]{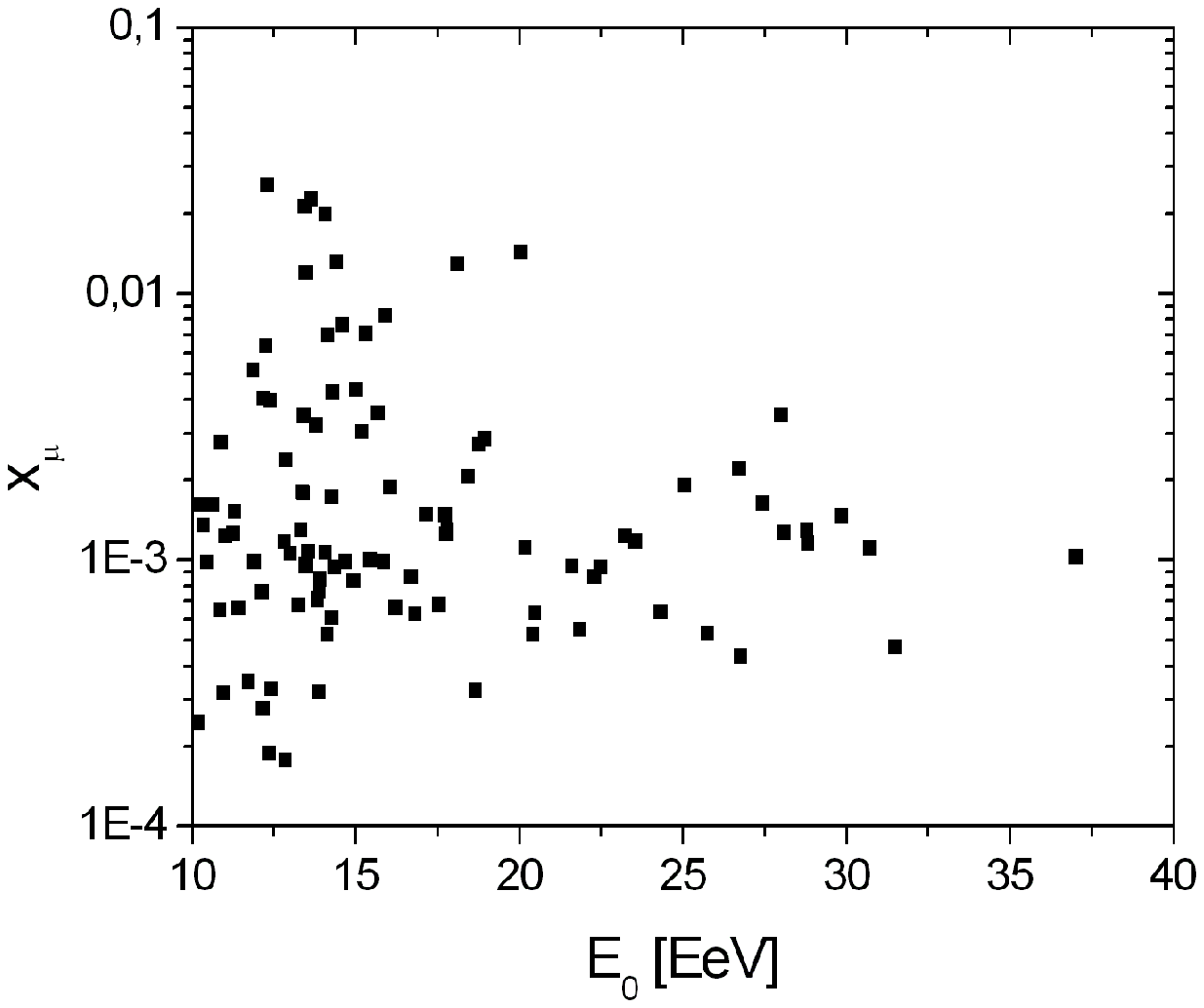}}
\caption[]{The ratio of the sum of the fluxes from the two secondary images in the energy range [0.9 $E_0$, $E_0$] to the flux of the source that would be received in absence of magnification effects in the same energy range ($\mu_{sec}$) (a) and the fraction $x_{\mu}(E_0)$ that would correspond to the required contribution of the source without lensing effects (b) as a function of the critical energy $E_0$ at which the multiple images appear.}
\end{center}
\end{figure}

An estimate of the maximum source density for which it should be possible to detect the events of the secondary images can be obtained by assuming equal intrinsic luminosity sources. We first determine for each simulated source the fraction $x(E_0)$ of the cosmic ray flux in the energy range [0.9 $E_0$, $E_0$] that corresponds to the events that are necessary to detect the caustic with chance probability smaller than $10^{-3}$ (plotted in Figure \ref{fig7b}). This fraction rises from $4 \times 10^{-3}$ at 10 EeV to around $10^{-2}$ at 30 EeV. The flux from the secondary images in the energy range [0.9 $E_0$, $E_0$] is largely amplified with respect to the one in the absence of lensing effects for most of the sources. Therefore, the required absolute luminosity of the source would be reduced by a factor $1/\mu_{sec}$, with $\mu_{sec}$ the ratio of the sum of the fluxes from the two secondary images in the energy range [0.9 $E_0$, $E_0$] to the flux of the source that would be received in the absence of magnification effects in the same energy range (shown in Figure \ref{fig8a} for each of the one hundred simulated sources as a function of their caustic energy). The fraction $x_{\mu}(E_0)=x(E_0) /\mu_{sec}(E_0)$ that would correspond to the required contribution of the source without lensing effects is plotted in Figure \ref{fig8b}. For the sources in the regions of the sky where the caustic magnification is more important, this fraction turns out to be smaller than $10^{-3}$. The prospects to detect the presence of caustics are best in these regions. From Figure \ref{fig8b} it can be seen that $x_{\mu}(E_0)$ is smaller than $10^{-3}$ for 41 of the 100 simulated sources. This corresponds roughly to a fraction $F=1/10$ of randomly chosen directions in the sky. The closest source in one of these regions is hence the best candidate to lead to a caustic detection.\\

Assuming an homogeneous distribution of sources with mean density $n$ and intrinsic luminosity $L$, we can estimate the mean flux expected from the closest source $\left <f_1 \right >$ as
\begin{equation}
\left <f_1 \right >=\int_0^{\infty} \frac{L}{4\pi r^2} \left (\exp({-\frac{4}{3} \pi F n r^3}) \right) \left (F n 4 \pi r^2 {\rm d}r\right )= \frac{\Gamma(1/3)}{6^{2/3} \pi^{1/3}} L \left (Fn \right)^{2/3},
\end{equation}
where the first parenthesis in the integral represents the probability of having no source within a distance $r$ in that sector of the sky and the second one represents the probability of having one source between $r$ and $r+{\rm d}r$.\\

On the other hand, the mean total flux from all sources in the sky located at a distance smaller than $R_{max}$ from Earth is given by
\begin{equation}
\left <f_{tot} \right >=LnR_{max}.
\end{equation}
Then, we can estimate the fraction of the mean flux from the closest source in a fraction $F$ of the sky to the total flux as
\begin{equation}
x_{\mu}=\frac{\left < f_1 \right >}{\left < f_{tot} \right>}= \frac{\Gamma(1/3)}{6^{2/3}\pi^{1/3}} \frac{F^{2/3}}{n^{1/3} R_{max}}.
\end{equation}

This allows to estimate an upper limit on $n$ given the fraction of the flux $x_{\mu}$ necessary to detect the signal from the secondary images and the fraction $F$ of the sky directions. For example, for $x_{\mu}=10^{-3}$ and $F=0.1$ and considering that cosmic rays of energy $E\simeq10$ EeV can arrive from $R_{max}\simeq 1$ Gpc, the upper limit on the source density for which it is expected to be possible to detect an excess due to secondary images of a source is $n\simeq 2 \times 10^{-3}$ Mpc$^{-3}$. This large value is compatible with the lower bounds that could be deduced from the searches of clustering at the highest energies.\\

This is a rough estimate of the mean density value with the simplifying assumption of equal intensity sources with an homogeneous distribution, so large fluctuations of this value can be expected. However, the order of magnitude of the obtained result implies that there are reasonable chances to detect this effect in the near future, at least if there is a nearby source of light ultra-high energy cosmic rays. For a source of heavy nuclei, the caustics will appear at energies $Z$ times larger and hence the observation of these effects will be much harder.\\

\section{Conclusions}

Magnetic lensing effects on ultra-high energy cosmic rays produce interesting phenomena which could be measured with the increasing statistics of present and future air shower experiments. A remarkable aspect is the appearance of multiple images of a source, which could be detectable due to the magnification of the flux near the critical energy at which the secondary images appear.\\

In this work, the typical angular distribution of secondary images was characterized with simulations of sources that present multiple images in a BSS-S model for the galactic magnetic field. Consequently, an algorithm to search for clustering of events of similar energy in angular stripes of $8^\circ \times 2^\circ$ in cosmic rays data was implemented. The significance of the excess is calculated with the Li \& Ma method and the probability of finding such a significance is determined from isotropic simulations.\\

We determined the minimum number of events that are required to detect an excess with probability smaller than $10^{-3}$ of occurring by chance in an isotropic distribution of arrival directions. We estimated also an upper limit to the source density which is needed to be able to detect such a number of events from a source. Our results indicate that such an excess should be detectable in the near future with the current ultra-high energy cosmic rays observatories if the source density is not too large and if there are nearby sources of light ultra-high energy cosmic rays. Detecting an excess due to the appearance of multiple images of a source would also allow to obtain information regarding the location of caustics which would be valuable in disentangling between different models for the galactic magnetic field.

\section{Acknowledgments}

This work is supported by ANPCyT (grant PICT 1334-06) and CONICET (grant PIP 01830).


\end{document}